\documentclass[12pt]{iopart}

\usepackage{iopams}
\def\dd{\displaystyle}
\usepackage{epsfig}
\begin{document}

\title{A Direct Approach to the Electromagnetic Casimir Energy in a Rectangular Waveguide}

\author{M.A. Valuyan, R. Moazzemi, and S.S. Gousheh}

\address{Department of Physics, Shahid Beheshti University, Evin, Tehran
19839, Iran}
\ead{m-valuyan@sbu.ac.ir}

\begin{abstract}
In this paper we compute the leading order Casimir energy for the
electromagnetic field (EM) in an open ended perfectly conducting rectangular waveguide in three spatial dimensions by a direct
approach. For this purpose we first  obtain the second quantized expression for the EM field with boundary conditions which would be appropriate for a waveguide. We then obtain the Casimir
energy by two different procedures.  Our main approach does
not contain any analytic continuation techniques. The second approach involves the routine
zeta function regularization along with some analytic
continuation techniques. Our two approaches yield identical results. This energy has
been calculated previously for the EM field in a rectangular
waveguide using an indirect approach invoking analogies between EM fields and
massless scalar fields, and using complicated analytic continuation techniques, and the results are identical to ours.  We have also
calculated the pressures on different sides and the total Casimir
energy per unit length, and plotted these quantities as a
function of the cross-sectional dimensions of the waveguide. We also
present a physical discussion about the rather peculiar effect of
the change in the sign of the pressures as a function of the shape
of the cross-sectional area.
\end{abstract}

\section{Introduction}
The Casimir effect is the physical manifestation of the change in
the zero point energy of a quantum field for different
configurations. The zero point configuration refers to one in which there does not exist any on-shell physical excitation of the field. The difference in the configurations could arise
either from the imposition of different boundary conditions on the
fields, or the presence of non-trivial spatial backgrounds
(\emph{e.g.} solitons). In 1948 Casimir predicted the existence of
this effect as an attractive force between two infinite parallel
uncharged perfectly conducting plates in vacuum \cite{h.b.g.}. This
effect was subsequently observed experimentally by Sparnaay in 1958
\cite{sparnaay} (for a general review on the Casimir effect, see
Refs.\,\cite{mostepanenko.,k.a.milton,milton.book}). Recently,
similar measurements have been done for other geometries, and their
precisions have been greatly improved \cite{
Lamoreaux.,Bressi.,Gusso.}. The manifestations of the Casimir effect
have been studied in many different areas of physics. For example,
the magnitude of the cosmological constant has been estimated using
the Casimir effect \cite{Elizalde.,bauer}. The effect has been also
studied within the context of string theory \cite{fabinger.}.
Recently this effect has been investigated in connection with the
properties of the space-time with extra dimensions
\cite{Poppenhaeger.}. The majority of the investigations related to
the Casimir effect concern the calculation of this energy or the
resulting forces for different fields in different geometries, such
as parallel plates \cite{h.b.g.}, cubes \cite{pra.47.4204.,
j.phys.35.2205., j.phys.34.11053.,prd.61.052110.,Lukosz.,ruggiero.},
cylinders \cite{j.opt.7.s86.,Mazzitelli.,Dalvit.}, and spherical
geometries \cite{prd.56.4896.,Bender.}. An interesting question is
the determination of the conditions under which the forces acting on
the boundaries for closed geometries are attractive or repulsive in
arbitrary spatial dimensions \cite{caruso.neto.431300.,
cavalcanti.69.065015.,Edery.,alnes.prd.74.105017., prl.95.250402.,
prd.56.2155.}. We should mention that in the calculations of the
Casimir energy many different regularization schemes or
renormalization programs have been used to remove the divergences,
and some of these techniques have been compared with each other
\cite{svaiter1.,svaiter2.,rodrigues.,p. wegrzyn.}. However, there are sometimes
ambiguities associated with the analytic continuation techniques.

Ambj{\o}rn and Wolfram \cite{wolf.} were the first to calculate the
Casimir energy in higher space-time dimensions, and in particular
derived an expression for the change in the vacuum energy due to a
rectangular box with p sides in d-dimensions for a massless and a
massive scalar field by summing the zero point energy of the eigenmodes. The divergences were
removed by using the following regularization and analytic
continuation procedures: Zeta function regularization, dimensional
regularization, and the reflection formula. The results are given in
terms of the Epstein zeta function\,\cite{epstein.,edery.}. Then, using analogies between EM  and massless scalar fields,
the Casimir energy for the EM radiation was indirectly deduced for the TE and TM modes in
a perfectly conducting rectangular box.

For the case of cavities, this energy has been calculated directly
for both the EM field and scalar fields. Usually two different methods are used. The more routine one involves the
aforementioned regularization and analytic continuation procedures.
The main ingredient of the second method is the subtraction of two
comparable configurations, sometimes supplemented by some
regularization procedures, such as the use of convergence factors.
The latter procedure was first used by Boyer for the calculation of
the the EM Casimir energy in a spherical cavity\,\cite{boyer.},
where he subtracted the zero point energies of two concentric
spheres, but with different sized inner cores. Analogous methods
were used for two parallel plates \cite{greiner} and rectangular
cavities \cite{Lukosz.}.

The primary purpose of this paper is to directly derive a closed form expression for the Casimir energy for the
EM field in a open-ended rectangular geometry in three spatial dimensions, which we shall henceforth call a waveguide. We do this by
directly finding the EM modes, quantizing the resulting field, and
calculating the zero point energy. We then obtain the Casimir energy using
first the usual program which involves complicated analytic
continuation techniques including zeta function and reflection formula, and second, a slight modification of the Boyer's
method, henceforth called the Box Subtraction Scheme (BSS). As we
shall show, in the latter procedure there is no need for any use of
analytic continuation techniques, and all of the divergences can be
removed without any ambiguities. Both of the results turn out to be
equivalent to those of \cite{wolf.}. Therefore, the secondary purpose of our work is to check the complicated analytic continuation techniques in common use. However, either of our direct
approaches to the problem has the advantage of being easily
extendable to higher orders in perturbation theory\cite{reza1.,reza2.}.

In the Section 2, we calculate the EM modes and the resulting zero
point energy in a rectangular waveguide. In Section 3, we first
calculate the Casimir energy using the zeta function regularization,
and then introduce the BSS to recalculate this energy. We then
compare our results with those of \cite{wolf.}. We then show that
our results for a rectangular waveguide of cross sectional area $a_1
\times a_2$ agrees with the established results for a cavity in the
appropriate limit. Then, we show that our results for the waveguide
reduces to those of the two infinite parallel plates, in the
appropriate limit. We then plot the Casimir energy per unit volume,
its contour plot per unit volume, and the Casimir energy per unit
length, all as a function of the cross-sectional lengths of the
waveguide. We then define and calculate the pressures on different
sides and plot them. We also present a physical discussion about the
rather peculiar effect of the change in the sign of one of the
pressures as a function of the shape of the cross-sectional area. We
summarize our results in Section 4.

\section{Zero point energy in a waveguide for the EM field}
The lagrangian density for the electromagnetic field is:
\begin{equation}\label{lagrangian den.}
  {\cal L} = \frac{1}{2}({\bf{E}}^2  - {\bf{B}}^2 ).
\end{equation}
For a conducting uncharged waveguide the electric and magnetic
fields can be easily written in terms of the vector potential for example in coulomb gauge\,(see
Ref.\,\cite{jackson}). Therefore to derive the EM modes, it is sufficient to calculate the vector
potential in the waveguide. However it is easier to find the
physical fields $\mathbf{E}$ and $\mathbf{B}$, and then calculate
the corresponding vector potential. This is due to the fact that it
is easier to impose the boundary conditions on the physical fields.
For this purpose, we need only compute $E_{z}$ for TM modes, and
$B_{z}$ for TE modes, where $z$ is defined to be the main axis of
the waveguide. The rest of the components can be calculated using
the Maxwell equations. Explicit expressions for all components of the EM fields are shown in\,\ref{Calculation1}.
Now, for each mode the
vector potential $\mathbf{A}_{\mathbf{k}}^{\lambda}(\mathbf{x},t)$ can be written in terms of the electric field as:
\begin{equation}\label{e5}
  {\bf{A}}_{\mathbf{k}}^\lambda  \left( {{\bf{x}},t} \right) =
  \frac{-ic}{\omega_{\mathbf{k}}} {\bf{E}}_{\mathbf{k}}^\lambda  \left(
  {{\bf{x}},t} \right),
\end{equation}
and its conjugate momenta is defined as:
\begin{equation}\label{cnojugate.mom.}
  {\mathbf{\Pi}}_{\mathbf{k}}^\lambda  \left( {{\bf{x}},t} \right) =
  \frac{1}{c^2} \frac{\partial
  }{{\partial t}}{\bf{A}}_{\mathbf{k}}^\lambda  \left( {{\bf{x}},t} \right),
\end{equation}
where $\lambda = \{ 1,\,2 \}$ indicate \{TE,\,TM\} modes,
respectively, and
\begin{equation}\label{e13}
  \omega _{\bf{k}}  = c|{\mathbf{k}}| = c\sqrt {\left( \frac{m \pi}{a_1}
  \right)^2  + \left( \frac{n \pi }{a_2 } \right)^2  + k_z^2
  }.
\end{equation}
Here $a_{1}$ and $a_2$ define the cross-sectional dimensions of the
rectangular waveguide. Since the set of all modes in the waveguide
are complete and orthonormal, one can expand the classical EM field
in terms of them \cite{sakurai},
\begin{eqnarray}\label{e6}\hspace{-2.3cm}
  \dd {\bf{A}} ({\bf{x}},t) =\dd\sum\limits_\lambda
  {\sum\limits_{m,n = 1}^\infty  \int  \frac{Ldk_z }{2\pi }
  \left[ {C^\lambda  \left( {\bf{k}} \right){\bf{A}}_{mn}^\lambda  \left( {{\bf{x}},t} \right)
  +c.c.} \right]}   \\
  \hspace{-1cm}
  \dd +\sum\limits_{m = 1}^\infty \int  {\frac{{Ldk_z }}{{2\pi }}}
  {\left[ {C^{TE} \left( {\bf{k}} \right){\bf{A}}_{m0}^{TE}
  \left( {{\bf{x}},t} \right) + c.c.} \right]}\dd+\sum\limits_{n = 1}^\infty\int{\frac{{Ldk_z }}{{2\pi }}}
  {\left[ {C^{TE} \left( {\bf{k}} \right){\bf{A}}_{0n}^{TE} \left( {{\bf{x}},t} \right) +
  c.c.}  \right]},\nonumber
\end{eqnarray}
where $C^\lambda(\mathbf{k})$ and $C^{\lambda *}(\mathbf{k})$ are
the expansion coefficients. We use the canonical quantization method
to quantize the field in the waveguide. For this purpose we let,
\begin{equation}\label{e7}
  \left\{ \begin{array}{ll}\vspace{.2cm}
  \dd C^\lambda  ({\bf{k}}) \to \sqrt {\frac{{\hbar c^2 }}{{2\omega _{\bf{k}} L}}}
  N^\lambda  \left( {\bf{k}} \right)a^\lambda  \left( {\bf{k}} \right), &  \\
  \dd C^{\lambda *} ({\bf{k}}) \to \sqrt {\frac{{\hbar c^2 }}{{2\omega _{\bf{k}} L}}}
  N^\lambda  \left( {\bf{k}} \right)a^{\lambda \dag } \left( {\bf{k}} \right) , \\
  \end{array} \right.
\end{equation}
where $a^{\lambda\dagger}(\mathbf{k})$ and $a^{\lambda}(\mathbf{k})$
are creation and annihilation operators. Now we impose the usual canonical commutation relations on the
fields and their conjugate momenta,
\begin{eqnarray}\label{e9}
  \left[ {{A}_i^\lambda  \left( {{\bf{x}},t} \right),{A}_j^{\lambda
  '} \left( {{\bf{x'}},t} \right)} \right] &=& \left[ {\Pi _i^\lambda
  \left( {{\bf{x}},t} \right),\Pi _j^{\lambda '} \left(
  {{\bf{x'}},t} \right)} \right] = 0,\nonumber\\\left[ {{A}_i^\lambda \left(
  {{\bf{x}},t} \right),\Pi _j^{\lambda '} \left( {{\bf{x'}},t}
  \right)} \right] &=& i\hbar \delta _{\lambda \lambda '} \delta _{ij}
  \delta \left( {{\bf{x}} - {\bf{x'}}} \right),
\end{eqnarray}
and this will result in the usual canonical commutation relations between the $a^{\lambda}(\mathbf{k})$ and $a^{\lambda\dagger}(\mathbf{k})$\,\cite{sakurai}. Using this relation along with Eq.\,(\ref{e7}) we can find the normalization
coefficients $N^{\lambda}(\mathbf{k})=\omega_{\mathbf{k}} / c $.
\par According to Eq.\,(\ref{lagrangian den.}), the hamiltonian is:
\begin{equation}\label{e11}
  H = \frac{1}{2}\int\limits_V d^3 x {\left( {\left| {{\bf{E}}\left(
  {{\bf{x}},t} \right)} \right|^2  + \left| {{\bf{B}}\left(
  {{\bf{x}},t} \right)} \right|^2 } \right)}.
\end{equation}
Integrating over the volume of the waveguide gives the energy, and
the vacuum expectation value of energy inside the waveguide is:
\begin{eqnarray}\label{e12}
  &&\hspace{-0.5cm}\dd\left\langle 0 \right|H\left| 0 \right\rangle =
  \dd\int {\frac{{Ldk_z }}{{2\pi }}}
  \left\{ {\sum\limits_{m,n = 1}^\infty  {{\hbar \omega _{\bf{k}} }}   +
  \sum\limits_{n = 1}^\infty  {\delta _{m0} \frac{{\hbar \omega _{\bf{k}} }}{2}}  +
  \sum\limits_{m = 1}^\infty  {\delta _{0n} } \frac{{\hbar \omega _{\bf{k}} }}{2}} \right\}
  \nonumber\\&&\hspace{-.6cm}
  \hspace{1.5cm}\dd = \int  {\frac{{Ldk_z }}{{2\pi }}} \sum\limits_{m,n = 0}^\infty
  {\left( {2 - \delta _{m0}  - \delta _{0n} } \right)}  \frac{{\hbar \omega _{\bf{k}}
  }}{2},
\end{eqnarray}
which simply means that the electromagnetic vacuum energy inside the
waveguide is the sum of the zero point energies of all possible
modes. This sum, and its analogues in any quantum field theory, always turn out to be infinite.

\section{The Casimir energy}

We obtained an expression for the zero point energy for the EM field
in a rectangular waveguide in the previous section. As mentioned earlier, the main purposes of this paper is to first obtain the Casimir energy using directly the EM field, and to avoid any use of analytic continuation techniques, utilizing BSS. Now we calculate
the resulting Casimir energy by two different methods. First is the
conventional method involving the usual regularization programs and
the ensuing analytic continuations, and second is our method. We shall find that these two methods
yield identical results for the leading order case, which agrees
with the results obtained indirectly in \cite{wolf.}.
\par The total energy of the vacuum for the EM field inside the
waveguide is given in Eq.\,(\ref{e12}). High frequency modes render
these sums formally divergent. Our first procedure involves the
usual zeta function regularization. That is, we shall compute the
following expression
\begin{equation}\label{zeta.reg.}
  E_{{\rm{Cas}}{\rm{.}}}  = \frac{\hbar }{2}\int {\frac{{Ldk_z
  }}{{2\pi }}} \sum\limits_{n = 0}^\infty  {\sum\limits_{m =
  0}^\infty {\left( {2 - \delta _{0n}  - \delta _{m0} } \right)} }
  \omega _{\bf{k}}^{d-2},
\end{equation}
which is convergent for $d<-1$. By calculating the integral in
Eq.\,(\ref{zeta.reg.}) and using the definition of the Epstein zeta
function \cite{epstein.} we obtain,
\begin{eqnarray}\label{eps.zeta.}
  \hspace{-1.5cm}E_{{\rm{Cas}}{\rm{.}}}  = \frac{{\hbar c^{d-2}L}}{{16}} \pi ^{d-
  \frac{3}{2}}\frac{{\Gamma \left( {\frac{{1 - d}}{2}}
  \right)}}{{\Gamma \left( {1-\frac{{d}}{2}} \right)}} \left[ {Z_2
  \left( {a_1^{ - 1} ,a_2^{ - 1} ;1 - d} \right) + Z_2 \left( {a_2^{ -
  1} ,a_1^{ - 1} ;1 - d} \right)} \right].
\end{eqnarray}
Note that the simple analytic continuation $d \rightarrow 3$ leads to divergent result!
However, we can use a simplified version of the reflection formula\,\cite{wolf.} applicable to this problem,
%General reflection formula is:
%\begin{eqnarray}\label{gen.reflect.formula.}\hspace{-1.5cm}\vspace{1cm}
%  \dd W_{p}\left(a_1 ,a_2 ,...,a_p ;s\right)&=& \dd \pi ^{ - s/2} \Gamma \left( {{\textstyle{s \over 2}}} \right)Z_p \left( {a_1 ,a_2 ,...,a_p ;s} %\right) \vspace{.3cm}\\  \dd&=& a_1^{ - 1} a_2^{ - 1} ...a_p^{ - 1} \pi ^{\left( {s - p} \right)/2} \Gamma \left( {{\textstyle{{p - s} \over 2}}} %\right)Z_p \left( {a_1^{-1} ,a_2^{-1} ,...,a_p^{ - 1} ;p - s} \right),\nonumber
%\end{eqnarray}
\begin{eqnarray}\label{gen.reflect.formula.}\hspace{-1.8cm}\vspace{1cm}
   \dd \pi ^{ \left(1-d\right)/2} \Gamma \left( {{\textstyle{1-d \over 2}}} \right)Z_2 \left( {a_1^{-1} ,a_2^{-1};1-d} \right) \vspace{.3cm}= a_1 a_2 \pi ^{-\left( {d+1} \right)/2} \Gamma \left( {{\textstyle{{d+1} \over 2}}} \right)Z_2 \left( {a_1 ,a_2;d+1} \right).
\end{eqnarray}
The analytic continuation embedded in the
reflection formula eliminates all of the infinities, and the final expression for
the Casimir energy in terms of the Epstein zeta function can be
written as\,\cite{epstein.}:
\begin{equation}\label{reflect.formula.}
  E_{{\rm{Cas}}{\rm{.}}}  = \frac{{ - \hbar cLa_1 a_2 }}{{32\pi ^2
  }}\left[ {Z_2 \left( {a_1 ,a_2;4} \right) + Z_2
  \left( {a_2 ,a_1 ;4} \right)} \right].
\end{equation}
\par
\begin{figure}[th] \hspace{4cm}\includegraphics[width=8cm]{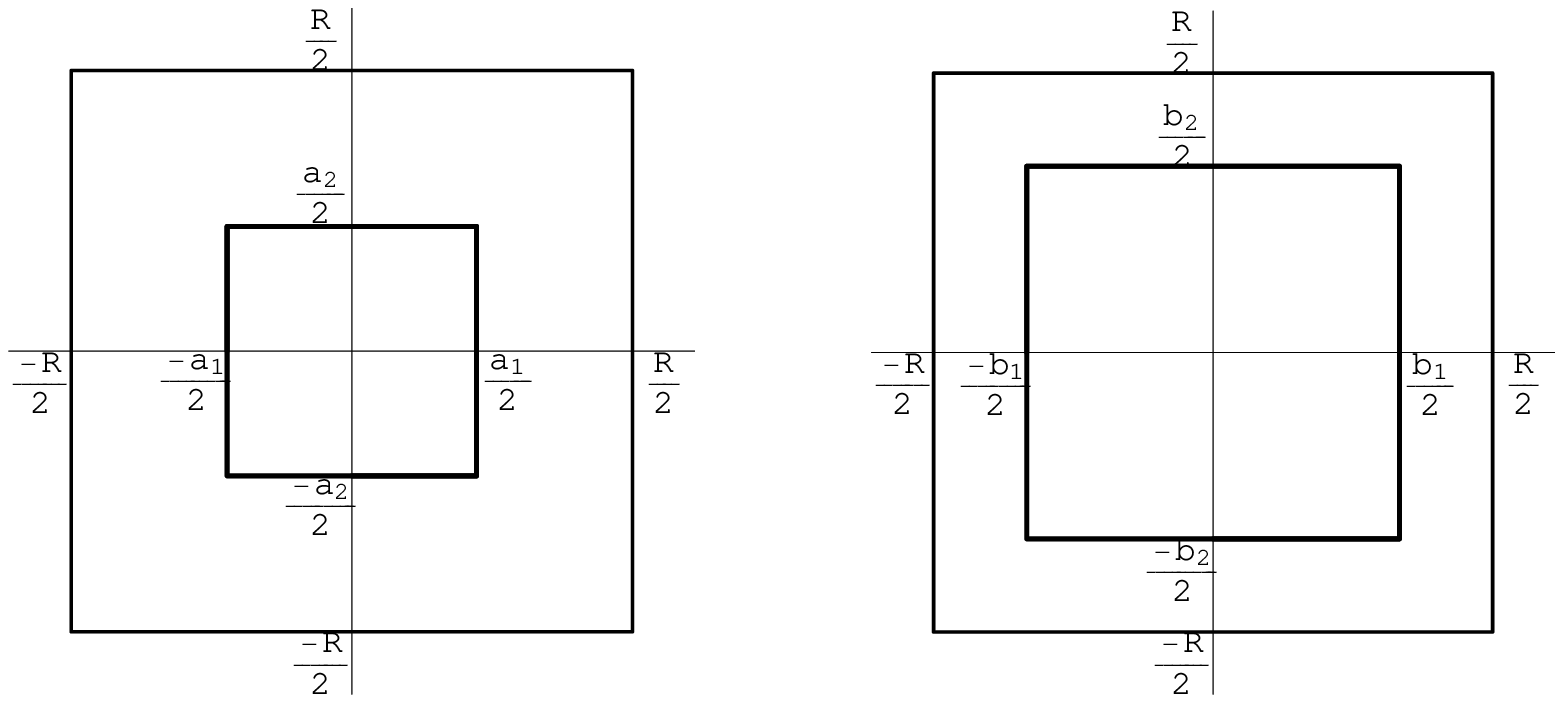}\caption{\label{fig.1} \small
  Left figure is ``$A$ configuration" and right one is ``$B$
  configuration".}
  \label{geometry}
\end{figure}

 As outlined above, the first method uses complicated analytic continuation techniques. We share the point of view with some authors that in general the use of analytic continuation techniques could lead to ambiguities \cite{kay.,mostepanenko.,reza1.,reza2.}. This ambiguity has already shown up in our calculations, as explicitly stated in the sentence below Eq.\,(\ref{eps.zeta.}). This is why we think it is worthwhile to obtain the Casimir energy by a method which does not use such techniques. As a first step towards this goal, we introduce two similar configurations, each of which consists of a waveguide enclosed in a larger one of cross-sectional area $R^2$, as shown
in Fig.\,(\ref{fig.1}). The Casimir energy can now be defined as:
\begin{equation}\label{e14}
  E_{{\rm{Cas.}}}  = \mathop {\lim }\limits_{b_1 /a,b_2 /a \to
  \infty } \left[ {\mathop {\lim }\limits_{R/b \to \infty } \left(
  {E_{A}  - E_{B} } \right)} \right],
\end{equation}
\begin{figure}[th] \hspace{4cm}\includegraphics[width=8cm]{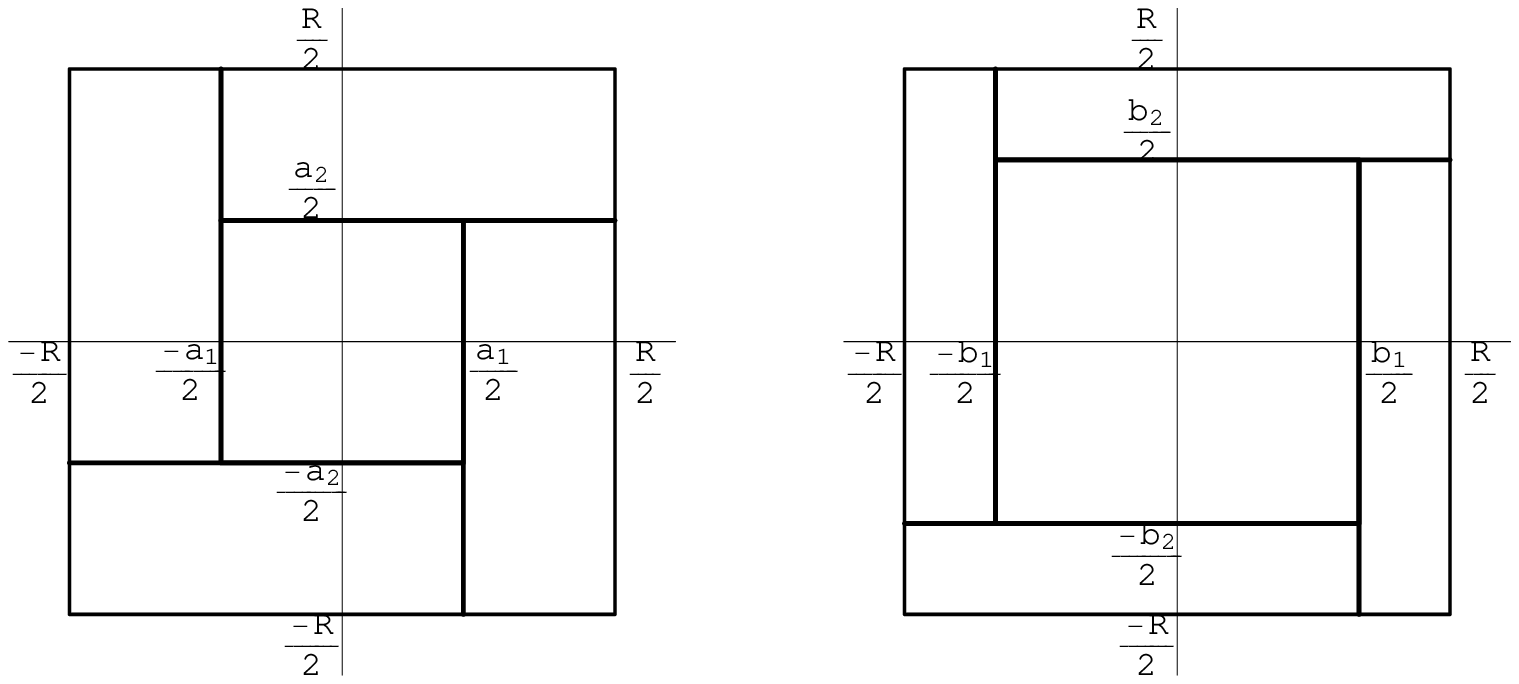}\caption{\label{fig.2} \small
  Left figure is ``$A'$ configuration" and right one is ``$B'$
  configuration".}
  \label{geometry}
\end{figure}
where $E_A$ ($E_B$) is the energy of configuration A (B), and
$a\equiv \mbox{Max}\{a_{1},a_{2}\}$ and $b\equiv
\mbox{Max}\{b_{1},b_{2}\}$. Subtraction of the zero point energy of B from A is
equivalent to the work done in deforming the configuration B to A.
Therefore, having chosen the same $R$ for both configurations, we expect this quantity per unit length to be finite on
physical grounds and to depend only on the dimensions of the original waveguide.
To calculate the Casimir  energy it is necessary to have an
expression for all the EM modes in the whole configuration. However,
calculation of the EM modes in the middle regions is very
cumbersome. Therefore, to simplify the task, without any loss of
generality, we define an alternative set of configurations in
Fig.\,(\ref{fig.2}). We can then define the Casimir energy as in
Eq.\,(\ref{e14}), but with following replacement $A\rightarrow A'$
and $B\rightarrow B'$. As we shall show explicitly in \ref{Calculation2}, as expected, all of the infinities automatically
cancel each other out upon subtracting the zero point energies of the two configurations, even
for finite values of $\{a_1,a_2,b_1,b_2,R\}$. It is important to note that for any finite value of $R$ the Casimir energies of the two sets of configurations depicted in Figs.(\ref{fig.1}) and (\ref{fig.2}) will differ by a finite amount, after the automatic cancelation of all the infinities. However, as we shall show explicitly in \ref{Calculation2}, the difference between the remaining finite terms due to the boundary waveguides  in the $\{A',B'\}$ configuration go to zero in the limit $R\rightarrow \infty$. Therefore the difference in the energies of the two sets of waveguides in the $\{A',B'\}$ configuration in the limit $R\rightarrow \infty$, is only due to
the difference between the two inner waveguides and this can be
properly defined to be the Casimir energy.

We can solve the
problem using BSS by first using the Abel-Plana Summation Formula
(APSF) \cite{Henrici} for the sums in our main equation for the zero
point energy, Eq.\,(\ref{e12}),
\begin{eqnarray}\label{e17}
  \dd\hspace{-2.5cm} E_{{\rm{Cas.}}}  &=&\dd \mathop{\lim}\limits_{\frac{b_1}{a},\frac{b_2}{a}\to
  \infty }\left[ {\mathop {\lim }\limits_{\frac{R}b \to \infty } \left( {E_{{\bf{A'}}}  - E_{{\bf{B'}}} } \right)} \right]  = \mathop {\lim }\limits_{\frac{b_1}{a},\frac{b_2}{a} \to \infty }
  \left\{ {\mathop {\lim }\limits_{\frac{R}{b} \to \infty } \frac{{\hbar c}}{2}\int  {\frac{{Ldk_z }}{{2\pi }}}
  \left[ {\sum\limits_{n = 1}^\infty  {g\left( n \right)} } \right]} \right\} \\ \hspace{-2.5cm}
  &=&\dd\mathop {\lim }\limits_{\frac{b_1}{a},\frac{b_2}{a} \to \infty }
  \left\{ {\mathop {\lim }\limits_{\frac{R}{b} \to \infty } \frac{{\hbar c}}{2}\int {\frac{{Ldk_z }}{{2\pi }}}
  \left[ {\frac{{ - 1}}{2}g\left( 0 \right) + \int_0^\infty  dx{g\left( x \right)} {\rm{ }} +
  i\int_0^\infty  dt{\frac{{g\left( {it} \right) - g\left( { - it} \right)}}{{e^{2\pi t}  - 1}}} {\rm{ }}} \right]}
  \right\},\nonumber
\end{eqnarray}
where $g\left( n \right)$ is defined in the \ref{Calculation2}. The
calculations for each term in the integrand is very lengthy, and are
done in the \ref{Calculation2}. However, here we like to briefly outline the calculations, especially the cancelation of infinities. The first term in the integrand contains infinite
terms some of which, as expected, cancel each other out due to the
BSS. However a few infinite terms remain. Similar cancelations occur
for the second term. The remaining infinite terms of these two terms
exactly cancel each other out. The third term (the branch-cut term) is
finite. Therefore all the infinities cancel each other out due to
the BSS, without resorting to any analytic continuation techniques, and more surprisingly, even without resorting to any regularization scheme.
The Casimir energy can be easily obtained by collecting all the finite pieces from the above three terms which are all of branch-cut types and obtained explicitly in the \ref{Calculation2}. The final expression for the Casimir energy, after taking the appropriate limits is,
\begin{eqnarray}\label{e18}
  \vspace{-1cm}&&\hspace{-2.3cm}\vspace{1cm}
  \dd E_{{\rm{Cas.}}}  = \dd\frac{{ - \hbar cL\zeta \left( 3 \right)}}{{32\pi}}
  \left(\frac{1}{a_1^2}+\frac{1}{a_2^2} \right)-\frac{{\hbar cL\pi ^2 a_2 a_1 }}{{1440 }}\left(\frac{1}{a_1^4}
  +\frac{1}{a_2^4}\right)
  \nonumber\\ &&\hspace{-2.4cm} \dd - \frac{{\hbar cL}}{16\pi}\sum\limits_{j = 1}^\infty
  {\frac{{\left( {e^{\frac{{2\pi ja_2 }}{{a_1 }}}  - 1} \right)a_1  +
  2\pi ja_2 e^{\frac{{2\pi ja_2 }}{{a_1 }}} }}{{ a_1 a_2^2 j^3\left( {e^{\frac{{2\pi ja_2 }}{{a_1 }}}  - 1}
  \right)^2 }}} - \frac{{\hbar cL}}{16\pi}\sum\limits_{j = 1}^\infty
  {\frac{{\left( {e^{\frac{{2\pi ja_1 }}{{a_2 }}}
  - 1} \right)a_2  +  2\pi ja_1 e^{\frac{{2\pi ja_1 }}{{a_2 }}} }}{{ a_2 a_1^2 j^3
  \left( {e^{\frac{{2\pi ja_1 }}{{a_2 }}}  - 1} \right)^2 }}}.
\end{eqnarray}
\begin{figure}[th] \hspace{4cm}\includegraphics[width=6.5cm]{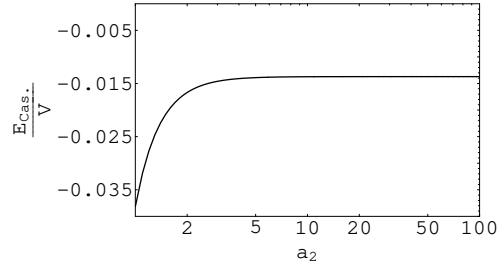}\caption{\label{fig.3} \small
  The Casimir energy, per unit volume, for the EM field inside
  a perfectly conducting rectangular waveguide
  in three spatial dimensions with cross-sectional area $a_{1}\times a_{2}$, in units $\hbar c=1$ and $a_1=1$.
  Note that the asymptotic value is the Casimir energy for two infinite parallel plates ($-\hbar c\pi ^2 /720$).
  When all the lengths are measured in units of $mm$ and the energy in $eV$, the factor for converting the energy density to
  $eV/mm^3$ is $1.978\times 10^{-4}$.}
  \label{geometry}
\end{figure}
\begin{figure}[th] \hspace{5cm}\includegraphics[width=5cm]{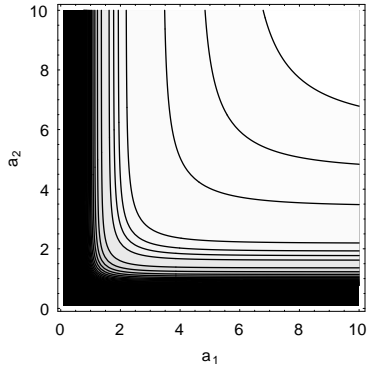}\caption{\label{fig.4} \small
  Contour plot of the Casimir energy, per unit volume, for the EM field inside a perfectly conducting rectangular
  waveguide, in units $\hbar c=1$.}
  \label{geometry}
\end{figure}
\par One can show that this expression for the Casimir energy is
identical to the previous one obtained by zeta function
regularization, i.e. Eq.\,(\ref{reflect.formula.}). One can also
easily show that the Casimir energy obtained here directly from the
second quantized form of the EM field for any waveguide in three
spatial dimensions, using any of our two programs, is identical with
the  results obtained indirectly in \cite{wolf.} using analogies
between massless scalar field and the EM field. As mentioned before,
the Casimir problem inside a rectangular cavity with perfectly
conducting walls has been solved directly and exactly, although the
final result is not in a closed form\,\cite{Lukosz.}. They have
computed the limit of their expression reducing to a waveguide with
square cross section. We have computed the limit of their results
for the slightly more general case of a waveguide with rectangular
cross section and the results are identical to ours. The other
extreme limit of the waveguide is when one of the sides approaches
infinity, for example $a_2$. Then, our result Eq.\,(\ref{e18}),
turns out to be exactly the Casimir energy for two infinite parallel
plates\,\cite{h.b.g.},
\begin{equation}\label{e25:E.bet.plats}
  E_{{\rm{Cas.}}}  =  - \frac{{\hbar c\pi ^2 La_2 }}{{720a_1^3 }}.
\end{equation}
The Casimir energy density is plotted in Fig.\,(\ref{fig.3}) and its
contour plot is illustrated in Fig.\,(\ref{fig.4}). Note that in
Fig.\,(\ref{fig.4}) the regions shaded darker correspond to lower
energies. \par We can define the Casimir pressure $P_{i}$ on the plate of the
waveguide whose perpendicular direction is $\hat{i}$, and its area
is $A_i$ by:
\begin{equation}\label{pressure}
  \dd P_{i}=-\frac{1}{A_{i}}\frac{\partial E_{\rm{Cas.}}}{\partial
  a_{i}} \qquad (i=1,2).
\end{equation}
\begin{figure}\hspace{4cm}\includegraphics[width=6.5cm]{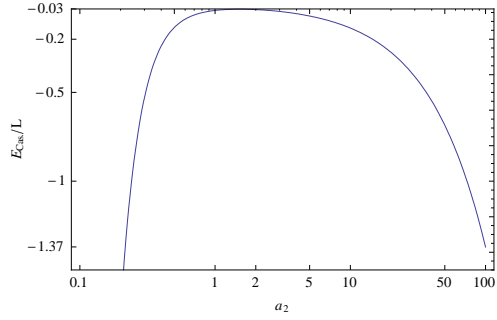}\caption{\label{tot.energy.}
\small
  The Casimir energy, per unit length, as a function of $a_2$
  for the EM field inside a conducting rectangular
  waveguide, in units $\hbar c=1$ and $a_1=1$. Note the existence of a maximum value signifies a change in the direction of the pressure.
  When all the lengths are measured in units of $mm$ and the energy in $eV$, the factor for converting the total energy, per unit length, to
  $eV/mm$ is $1.978\times 10^{-4}$.}
  \label{geometry}
\end{figure}
\begin{figure}[th] \hspace{4cm}\includegraphics[width=6.5cm]{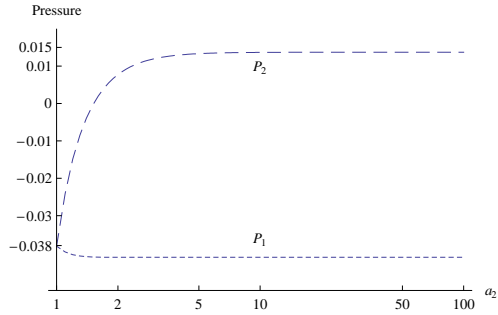}\caption{\label{pressure.} \small
  The Casimir pressure as a function of $a_2$ for the EM field inside a conducting rectangular
  waveguide, in units $\hbar c=1$ and $a_1=1$. Note $P_2$ increases to a positive value after the critical value $a_2 \approx1.52$.
  When all the lengths are measured in $mm$, the factor for converting the pressure to
  $\mu Pa$ is $3.166\times 10^{-8}$.}
  \label{geometry}
\end{figure}
In Fig.\,(\ref{tot.energy.}) we display the Casimir energy, per unit
length, as a function of $a_2$ for $a_1=1$. Note the existence of a
maximum indicates that as $a_2$ increases for fixed $a_1$, the
pressure $P_2$ increases from a negative (inwards) to a positive
value (outwards), while $P_1$ approaches the value appropriate for
the parallel plate problem (Fig.\,\ref{pressure.}). Analogous
findings have been reported for the rectangular cavity problem
\cite{pra.47.4204.,Lukosz.,prd.56.2155.}. We have come up with the following
physical reasoning for this phenomena: As $a_2$ increases, for fixed
$a_1$, past a critical threshold of about $1.52$, the total energy
starts to decrease although, any two opposite pairs of  sides always
attract each other. That is, with any further increase in $a_2$, the
increase in the energy due to the attraction of the sides labeled
$A_2=a_1\times L$ is more than compensated by the reduction in the
energy due to the attraction of the sides labeled $A_1=a_2\times L$.

\section{Conclusion}

In this paper, we have obtained the Casimir energy for EM
field in a rectangular waveguide by directly finding the EM modes
inside the waveguide, and summing over the zero modes of the
corresponding second quantized EM field operator. We compute the
sums using two completely different methods. First is the zeta
function analytic continuation method. Second, is the BSS whose
basis is confining the waveguide inside a large one and computing
the difference in the vacuum energies in two comparable configurations. The results of these two techniques are identical.
However, the latter provides a mechanism by which precise
cancelations of divergences occur without using any analytic
continuation or regularization schemes. Our results also turn out to be identical to those of
\cite{Lukosz.,wolf.} in the appropriate limit. However in reference
\cite{wolf.} the results were obtained indirectly by using some
analogies between the EM field and massless scalar fields. The
direct method has the added advantage of being easily extendable to
calculations of radiative corrections to the Casimir energy. We have also
computed and displayed the results for the Casimir energy, its
density, and pressures as a function of the cross sectional sides of
the waveguide. The physical reasons for the phenomena of the change
in the sign of one of the pressures are also given.

\appendix
\section{Electromagnetic Modes in Waveguides} \label{Calculation1}
In this Appendix we present explicit formulae for the TE and TM modes inside a rectangular waveguide with inner dimensions $a_1$ and $a_2$\,\cite{jackson}.
For the TE waves the wave equation for $B_{z}$ is:
\begin{eqnarray}\label{app.a.TEwave.eq.}
\nabla _t^2 B_z ( \mathbf{x},t) + \mathbf{k}_t^2 B_z (\mathbf{x},t) = 0,
\end{eqnarray}
where $\mathbf{k}_{t}$ is transverse momentum. The solution for $B_{z}$ satisfying the appropriate boundary conditions is,
\begin{eqnarray}\label{app.a.TEwavesol.eq.}
B_z  = B_{0z} \cos \left( {\frac{{n\pi }}{{a_1 }}\left( {x + \frac{{a_1 }}{2}} \right)} \right)\cos \left( {\frac{{m\pi }}{{a_2 }}\left( {y + \frac{{a_2 }}{2}} \right)} \right)
\end{eqnarray}
and $\mathbf{k}_t^2  = \left( {{\textstyle{{n\pi } \over {a_1 }}}} \right)^2  + \left( {{\textstyle{{m\pi } \over {a_2 }}}} \right)^2$. Now, we can use the Maxwell equations to obtain all the components of the EM fields,
\begin{eqnarray}\label{app.a.TEmodes.eq.}
\begin{array}{l}\vspace{.2cm}
 \dd{\bf{B}}_{nm} ({\bf{x}},t) = B_{0z}\Bigg\{ \frac{{ik_z }}{{\mathbf{k}_t^2 }} \Bigg[\frac{{ - n\pi }}{{a_1 }}\sin \left(\frac{{n\pi }}{{a_1 }}(x + \frac{{a_1 }}{2})\right)\cos \left(\frac{{m\pi }}{{a_2 }}(y + \frac{{a_2 }}{2})\right)\hat \mathbf{x} \\ \vspace{.2cm}\dd
  - \frac{{m\pi }}{{a_2 }}\cos \left(\frac{{n\pi }}{{a_1 }}(x + \frac{{a_1 }}{2})\right)\sin \left(\frac{{m\pi }}{{a_2 }}(y + \frac{{a_2 }}{2})\right)\hat \mathbf{y}\Bigg] \\ \vspace{.2cm}\dd
  +  \cos \left(\frac{{n\pi }}{{a_1 }}(x + \frac{{a_1 }}{2})\right)\cos \left(\frac{{m\pi }}{{a_2 }}(y + \frac{{a_2 }}{2})\right)\hat \mathbf{z}\Bigg\} e^{ik_z z - i\omega _{\bf{k}} t}  \\ \\ \vspace{.2cm} \dd
 {\bf{E}}_{nm} ({\bf{x}},t) = \frac{{i\omega _{\bf{k}} }}{{\mathbf{k}_t^2 }c}B_{0z} \Bigg[-\frac{{m\pi }}{{a_2 }}\cos \left(\frac{{n\pi }}{{a_1 }}(x + \frac{{a_1 }}{2})\right)\sin \left(\frac{{m\pi }}{{a_2 }}(y + \frac{{a_2 }}{2})\right)\hat \mathbf{x} \\ \vspace{.2cm} \dd+\frac{{n\pi }}{{a_1 }}\sin \left(\frac{{n\pi }}{{a_1 }}(x + \frac{{a_1 }}{2})\right)\cos \left(\frac{{m\pi }}{{a_2 }}(y + \frac{{a_2 }}{2})\right)\hat \mathbf{y} \Bigg]e^{ik_z z - i\omega _{\bf{k}} t}
 \end{array}
\end{eqnarray}
For the TM waves the wave equation for $E_{z}$ is:
\begin{eqnarray}\label{app.a.TMwave.eq.}
\nabla _t^2 E_z ( \mathbf{x},t) + \mathbf{k}_t^2 E_z (\mathbf{x},t) = 0,
\end{eqnarray}
where $\mathbf{k}_{t}$ is transverse momentum. The solution for $E_{z}$ satisfying the appropriate boundary conditions is,
\begin{eqnarray}\label{app.a.TMwavesol.eq.}
E_z  = E_{0z} \sin \left( {\frac{{n\pi }}{{a_1 }}(x + \frac{{a_1 }}{2})} \right)\sin \left( {\frac{{m\pi }}{{a_2 }}(y + \frac{{a_2 }}{2})} \right),
\end{eqnarray}
Now, we can use the Maxwell equations to obtain all the components of the EM fields,
\begin{eqnarray}\label{app.a.TMmodes1.eq.}
 \begin{array}{l}\vspace{.2cm}\hspace{-2.5cm}
 \dd {\bf{B}}_{nm} ({\bf{x}},t) = \frac{{i\omega_{\mathbf{k}} }}{{\mathbf{k}_t^2}}E_{0z} \Bigg[\frac{{n\pi }}{{a_1 }}\cos \left(\frac{{n\pi }}{{a_1 }}(x + \frac{{a_1 }}{2})\right)\sin \left(\frac{{m\pi }}{{a_2 }}(y + \frac{{a_2 }}{2})\right)\hat \mathbf{y}  \\ \vspace{.2cm}\dd
 - \frac{{m\pi }}{{a_2 }}\sin \left(\frac{{n\pi }}{{a_1 }}(x + \frac{{a_1 }}{2})\right)\cos \left(\frac{{m\pi }}{{a_2 }}(y + \frac{{a_2 }}{2})\right)\hat \mathbf{x}\Bigg]e^{ik_z z - i\omega _{\bf{k}} t}
 \end{array}
\end{eqnarray}
\begin{eqnarray}\label{app.a.TMmodes2.eq.}
 \begin{array}{l} \vspace{.2cm}\hspace{-2.5cm}\dd
 {\bf{E}}_{nm} ({\bf{x}},t) = \Bigg\{ \frac{{ik_{z}}}{{\mathbf{k}_t^2 }c}E_{0z} \Bigg[\frac{{n\pi }}{{a_1 }}\cos \left(\frac{{n\pi }}{{a_1 }}(x + \frac{{a_1 }}{2})\right)\sin \left(\frac{{m\pi }}{{a_2 }}(y + \frac{{a_2 }}{2})\right)\hat \mathbf{x} \\ \vspace{.2cm}\hspace{-.5cm}\dd
 +\frac{{m\pi }}{{a_2 }}\sin \left(\frac{{n\pi }}{{a_1 }}(x + \frac{{a_1 }}{2})\right)\cos \left(\frac{{m\pi }}{{a_2 }}(y + \frac{{a_2 }}{2})\right)\hat \mathbf{y}\Bigg]  \\ \vspace{.2cm}\hspace{-.5cm}\dd
 +E_{0z} \sin \left(\frac{{n\pi }}{{a_1 }}(x + \frac{{a_1 }}{2})\right)\sin \left(\frac{{m\pi }}{{a_2 }}(y + \frac{{a_2 }}{2})\right)\hat \mathbf{z}\Bigg\} e^{ik_z z - i\omega _{\bf{k}} t}
 \end{array}
\end{eqnarray}

\section{Explicit Calculations of the Casimir Energy} \label{Calculation2}

\setcounter{equation}{0}
\renewcommand{\theequation}{\Alph{section}.\arabic{equation}}
In this Appendix we show the main steps necessary to derive the main expression for the Casimir energy for a perfectly conducting waveguide with cross sectional area $a_1 \times a_2$, starting with its original definition, Eq.\,(\ref{e17}), and using the second set of configurations in fig.\,(\ref{fig.2}). Then the Casimir energy can be written as:
\begin{eqnarray}\label{1}
 \hspace{-2cm}E_{{\rm{Cas}}{\rm{.}}}  = \left[ {E_0 \left( {a_1 ,a_2 } \right) + 2E_0 \left( {\frac{{R - a_1 }}{2},\frac{{R + a_2 }}{2}} \right) + 2E_0 \left( {\frac{{R + a_1 }}{2},\frac{{R - a_2 }}{2}} \right)} \right]\vspace{5mm}\hspace{-2cm} \nonumber\\ - \Bigg\{ {a_{1} \to b_{1}},{a_{2} \to b_{2}} \Bigg\}.
\end{eqnarray}
The main ingredient of Eq.\,(\ref{e17}) for the Casimir energy is its integrand whose explicit form can be obtained from Eq.\,(\ref{1}),
\begin{equation}\label{ea.g(n)}
  \begin{array}{ll}\vspace{.2cm}\hspace{-2.5cm}
  \dd \sum\limits_{n=1}^\infty g\left( n \right) = \sum\limits_{n=1}^\infty \left\{ S_n \left( {a_1 } \right) + S_n \left( {a_2 } \right)+
  2{\sum\limits_{m = 1}^\infty  {S_{m,n} \left( {a_1 ,a_2 } \right)} } \right.
  \\ \hspace{-2.cm}
  \dd\vspace{.2cm}  \hspace{.5cm}+2 \left[ S_n \left( {\frac{{R - a_1 }}{2}} \right) + S_n \left( {\frac{{R + a_2 }}{2}} \right)
  +2\sum\limits_{m = 1}^\infty  {S_{m,n} \left( {\frac{{R - a_1 }}{2},\frac{{R + a_2 }}{2}}
  \right)} \right]
  \\ \hspace{-2.cm}
  \left. \dd\vspace{.2cm} \hspace{.5cm}+2 \left[ S_n \left( {\frac{{R + a_1 }}{2}}
  \right) + S_n \left( {\frac{{R - a_2 }}{2}} \right)
  +2\sum\limits_{m = 1}^\infty  {S_{m,n} \left( {\frac{{R + a_1 }}{2},\frac{{R - a_2 }}{2}}
  \right)} \right] \right\}
  \vspace{0.2cm}\\ \hspace{-1.3cm} - \Bigg\{  a_{1}\rightarrow b_{1}, a_{2}\rightarrow
  b_{2} \Bigg\} ,
  \end{array}
\end{equation}
where,
\begin{equation}\label{ea2}
  \begin{array}{c}\hspace{-1.5cm}
  \dd S_{m,n} (x,y) = \dd\sqrt {\left( {\frac{{m\pi }}{x}} \right)^2  + \left( {\frac{{n\pi }}{y}} \right)^2  + k_z^2
  }, \hspace{5mm} \mbox{and} \hspace{5mm}
  \dd S_n (x) = \dd\sqrt {\left( {\frac{{n\pi }}{x}} \right)^2  + k_z^2
  },
  \end{array}
\end{equation}
where $x$ and $y$ denote the dimensions of the inner boxes. As we shall show explicitly below, the BSS, symbolically indicated in the last line of Eq.\,(\ref{ea.g(n)}), renders that quantity finite. Equation\,(\ref{ea.g(n)})  is manifestly symmetric in all its double arguments symbolically denoted by $x$ and $y$, and the calculations will greatly simplify if we make use of this symmetry in the following form,
\begin{equation}\label{ea3}
  \sum\limits_{n = 1}^\infty  {\sum\limits_{m = 1}^\infty  {S_{m,n}
  (x ,y )} }  = \sum\limits_{n = 1}^\infty  {\sum\limits_{m =
  1}^\infty  {\frac{1}{2}\left( {S_{m,n} (x ,y ) + S_{m,n} (y
  ,x )} \right)} }.
\end{equation}
One of the main tools that we use to obtain the final results is the Abel-Plana Summation Formula (APSF),
\begin{equation}\label{Able-Plana}
\dd \sum\limits_{n=1}^\infty g\left( n \right)={\frac{{ - 1}}{2}g\left( 0 \right) + \int_0^\infty  dx{g\left( x \right)} {\rm{ }} +
  i\int_0^\infty  dt{\frac{{g\left( {it} \right) - g\left( { - it} \right)}}{{e^{2\pi t}  - 1}}} {\rm{ }}}
\end{equation}
The last term in the APSF is called the branch-cut term and always leads to finite results. Now we use the APSF for Eq.\,(\ref{ea.g(n)}). The first term gives,
\begin{eqnarray}\label{ea1}
  \hspace{-1.5cm}\dd \frac{-1}{2}g\left( 0 \right) = \dd\frac{-1}{2}\Bigg\{k_{z}+k_{z}+ \sum\limits_{m =
  1}^\infty S_m \left( {a_1 } \right)+\sum\limits_{m =
  1}^\infty S_m \left( {a_2 } \right)\\+2\Bigg[k_{z}+k_{z}+\sum\limits_{m =
  1}^\infty S_m \left( {\frac{{R - a_1 }}{2}}
  \right)+\sum\limits_{m = 1}^\infty S_m \left( {\frac{{R + a_2}}{2}} \right)\Bigg]\nonumber\\+2\Bigg[ k_{z}+k_{z}+\sum\limits_{m = 1}^\infty S_m \left( {\frac{{R + a_1 }}{2}} \right)  +\sum\limits_{m = 1}^\infty S_m \left( {\frac{{R - a_2 }}{2}} \right)\Bigg]\nonumber\\-\mbox{\large\{} a_{1}\rightarrow b_{1}, a_{2}\rightarrow b_{2}\large\}\Bigg\}.\nonumber
\end{eqnarray}
Note that each of the $k_z$ terms give quadratically divergent terms when integrated. However, BSS cancel all these terms between $A'$ and $B'$ configurations. Now we can use the APSF again for the remaining sums. To simplify the expression, we use the following change of variables,
\begin{equation}  \label{ea4}
  \int_0^\infty  dm{\sqrt {\left( {\frac{{m\pi }}{x}} \right)^2  +
  k_z^2 } }\qquad {\qquad{{t =
  \frac{m\pi}{x}}\qquad}\over{}}{\hspace{-.15cm}\tiny{\succ}}\qquad\frac{x}{\pi
  }\int_0^\infty dt{\sqrt {t^2 + k_z^2 } }.
\end{equation}
The result is:
\begin{eqnarray}\label{ea065}
   \hspace{-2.5cm}\vspace{.2cm}
  \dd\frac{- 1}{2}g\left( 0 \right) =\dd \frac{-1}{2}\Bigg\{ -k_{z}+\frac{a_1 + a_2 }{\pi }\int_0^\infty  dt{\sqrt {t^2  + k_z^2 }}
  + B_{k_z } \left( {a_1 } \right) + B_{k_z } \left( a_2  \right)\nonumber\\ \hspace{-10mm}+ 2\Bigg[-k_{z}+\left(\frac{R-a_{1}}{2\pi}+\frac{R+a_{2}}{2\pi}\right)\int_0^\infty  dt{\sqrt {t^2  + k_z^2 }}+B_{k_z} \left( \frac{R - a_1}{2} \right)+B_{k_z} \left( \frac{R + a_2 }{2}\right)\Bigg]\nonumber
  \\ \hspace{-10mm}+ 2\Bigg[-k_{z}+\left(\frac{R+a_{1}}{2\pi}+\frac{R-a_{2}}{2\pi}\right)\int_0^\infty  dt{\sqrt {t^2  + k_z^2 }}+B_{k_z} \left( \frac{R + a_1}{2} \right)+B_{k_z} \left( \frac{R - a_2 }{2}\right)\Bigg]\nonumber\\ \hspace{-10mm}- \mbox{\large\{} a_{1}\rightarrow b_{1}, a_{2}\rightarrow
  b_{2} \mbox{\large\}}\Bigg\}.
\end{eqnarray}
Where the branch-cut terms in this case become,
\begin{equation}\label{ea6}
  B_{k_z } \left( x \right) =  - 2\int_{\frac{{k_z x}}{\pi }}^\infty dt
  {\frac{{\sqrt {\left( {\frac{{t\pi }}{x}} \right)^2  - k_z^2 }
  }}{{e^{2\pi t}  - 1}}}.
\end{equation}
As is apparent from this expression we have terms which again lead to divergent results when integrated, or are directly divergent. They include the previous $k_z$ terms along with new ones involving square roots. Almost all of these terms cancel due to our BSS, with only the first square root term remaining. As we shall show, it will exactly cancel the analogous terms coming from the second term of APSF, Eq.\,(\ref{Able-Plana}).
The remaining terms are,
\begin{eqnarray}\label{ea5}
  \hspace{-2.5cm}\vspace{.2cm}
  \dd\frac{- 1}{2}g\left( 0 \right) =\dd \frac{-1}{2}\left[\Bigg\{ \frac{a_1 + a_2 }{\pi }\int_0^\infty  dt{\sqrt {t^2  + k_z^2 }}
  + B_{k_z } \left( {a_1 } \right) + B_{k_z } \left( a_2  \right)+ 2B_{k_z} \left( \frac{R - a_1}{2} \right) \right.
  \\ \hspace{-15mm} \left.
  \dd\vspace{.2cm}+2B_{k_z} \left( \frac{R + a_2 }{2}\right)+ 2B_{k_z }
  \left( \frac{R + a_1 }{2} \right) + 2B_{k_z } \left( {\frac{{R - a_2 }}{2}} \right)\Bigg\}- \mbox{\large\{} a_{1}\rightarrow b_{1}, a_{2}\rightarrow
  b_{2} \mbox{\large\}}\right].\nonumber
\end{eqnarray}
The second term of APSF gives,
\begin{eqnarray}\label{app.b.integ.term.}
  \vspace{.4cm}\hspace{-2.5cm}
  \vspace{2mm}\dd \int_0^\infty  dx {g\left( x \right)} =\Bigg\{ \frac{{a_2 }}{\pi } \sum\limits_{m = 1}^\infty \int_0^\infty  {dt\sqrt {\left( {{\textstyle{{m\pi } \over {a_1 }}}} \right)^2  + t^2  + k_z^2 } }
  +  {\frac{{ a_2 }}{\pi }}\int_0^\infty  {dt\sqrt {t^2  + k_z^2 } }  \\ \vspace{2mm}\hspace{-2.3cm} \dd+ 2\Bigg[ {\frac{{R + a_2 }}{{2\pi }}\sum\limits_{m = 1}^\infty  {\int_0^\infty  {dt\sqrt {\left( {{\textstyle{{2m\pi } \over {R - a_1 }}}} \right)^2  + t^2  + k_z^2 } } } }   + \frac{{R - a_1 }}{{2\pi }}\sum\limits_{m = 1}^\infty  {\int_0^\infty  {dt\sqrt {t^2  + \left( {{\textstyle{{2m\pi } \over {R + a_2 }}}} \right)^2  + k_z^2 } } }\nonumber   \\ \vspace{2mm}\hspace{-2.3cm} \dd
  + \left( {\frac{{R - a_1 }}{{2\pi }} + \frac{{R + a_2 }}{{2\pi }}} \right)\int_0^\infty  {dt\sqrt {t^2  + k_z^2 } }  \Bigg]
  + \large\{a_{1} \to a_{2},a_{2} \to a_{1}\large\}\Bigg\} - \Bigg\{ {a_1  \to b_1 ,a_2  \to b_2 } \Bigg\},\nonumber
\end{eqnarray}
Carrying out analogous steps as for the first term of APSF, we obtain:
\begin{eqnarray}\label{ea7}
  \vspace{.4cm}\hspace{-2.3cm}
  \dd\int_0^\infty  dx {g\left( x \right)}=\dd \Bigg\{ \frac{a_1 + a_2}{2\pi}\int_0^\infty
  dt\sqrt {t^2  + k_z^2 } + \int_0^\infty  dt \left[ \frac{a_1}{\pi }B_{k_z ,t} \left( {a_2} \right) +
  \frac{a_2}{\pi}B_{k_z ,t} \left( {a_1} \right) \right.\nonumber
  \\ \left. \hspace{-1.5cm}  \dd\vspace{.4cm}
  + \frac{R + a_2 }{\pi }B_{k_z ,t} \left( \frac{R - a_1 }{2} \right) +
  \frac{{R - a_1 }}{\pi }B_{k_z ,t} \left( \frac{R + a_2 }{2} \right) +
  \frac{{R - a_2 }}{\pi }B_{k_z ,t} \left( \frac{R + a_1 }{2}
  \right)\right. \vspace{.3cm} \nonumber \\ \left.\hspace{-1.5cm}  \dd
  + \frac{R + a_1 }{\pi }B_{k_z ,t} \left( \frac{R - a_2 }{2} \right)
  \right]\Bigg\} -\Bigg\{ a_{1}\rightarrow b_{1}, a_{2}\rightarrow
  b_{2} \Bigg\},
\end{eqnarray}
where the branch-cut term in this case becomes,
\begin{equation}\label{ea8}
  B_{k_z ,t} \left( x \right) =  - 2\int_{\frac{{x\sqrt {t^2  + k_z^2
  } }}{\pi }}^\infty  {dp\frac{{\sqrt {\left( {\frac{{\pi
  p}}{x}} \right)^2  - \left( {t^2  + k_z^2 } \right)} }}{{e^{2\pi p}
  - 1}}}.
\end{equation}
Note that by summing the first two terms of APSF (Eqs.\,(\ref{ea5}) and (\ref{ea7})),
the divergences cancel each other out exactly, and no infinite term remains. This is solely due the fact that in the BSS we have chosen the cross-sectional area of the outer waveguides to be equal for both $A'$ and $B'$ configurations. The third term of the APSF which contains only two kinds of branch-cut terms. The first type is given in Eq.\,(\ref{ea6}) and the second type is,
\begin{equation}\label{ea10}
  B\left( {y,S_m \left( x \right)} \right) =  - 2\int_{\frac{{yS_m
  \left( x \right)}}{\pi }}^\infty dt{\frac{{\sqrt {\left(
  {\frac{{t\pi }}{y}} \right)^2  - S_m^2 \left( x \right)}
  }}{{e^{2\pi
  t}  - 1}}} {\rm{ }}.
\end{equation}
Note that only finite terms of branch-cut types remain from all three terms of the APSF. Now we just have to collect them all and integrate them over $k_z$ as indicated in Eq.\,(\ref{e17}). The integral of the three different kind of branch cuts can be easily calculated and
the results for each are:
\begin{eqnarray}\label{branch.1}
  \int {\frac{{Ldk_z }}{{2\pi }}} B_{k_z } \left( x \right) = \frac{{
  - L\zeta \left( 3 \right)}}{{8\pi x^2 }},
\end{eqnarray}
\begin{eqnarray} \label{branch.2}
  &\dd\int& {\frac{{Ldk_z }}{{2\pi }}} \left[ {\int {dt} B_{k_z ,t} \left(
  x \right)} \right] = \frac{{ - L\pi ^3 }}{{720x^3 }},
\end{eqnarray}
\begin{eqnarray} \label{branch.3}&
 \vspace{.2cm} \dd\int &{\frac{{Ldk_z }}{{2\pi }}}  {\sum\limits_{m = 1}^\infty
  {B \left( {y,S_m \left( x \right)} \right)} }
  =\frac{{ - L}}{{8\pi }}\sum\limits_{j = 1}^\infty  {\frac{{x\left(
  {e^{\frac{{2\pi jy}}{x}} }-1\right) + 2\pi j y e^{^{\frac{{2\pi jy}}{x}}
  } }}{{xy^2 j^3 \pi ^2 \left( {e^{\frac{{2\pi jy}}{x}}  - 1}
  \right)^2 }}}.
\end{eqnarray}
The sum over $j$ in the last equation comes from the Taylor
expansion of the denominator of that particular branch-cut term Eq.\,(\ref{ea10}). An extremely important point to mention is that the remaining finite contribution to the Casimir energy coming from the outer waveguides, even after BSS, is nonzero for finite values of the dimensions of the waveguides. However, as we shall show below, in the limit of large $R$, there is partial cancelation between those terms, and the remaining terms go to zero in the limit $R \rightarrow \infty$. This shows that the outer waveguides have done their job in the BSS of canceling infinities, without leaving any finite contribution to the Casimir energy in the limit $R \rightarrow \infty$. The final step
in calculation of the Casimir energy is a calculation of the limits
in the appropriate order as indicated in Eq.\,(\ref{e17}). Then the
final result is given in Eq.\,(\ref{e18}).

\par The finite contribution to the Casimir energy coming from the outer waveguides is,
\begin{eqnarray}\label{ea3-1}
  \hspace{-1.1cm}\vspace{1cm}
  \dd E_{{\rm{Cas.}}}^{{\rm{Outer}}}  =2\Bigg\{ \dd\frac{{ - \hbar cL\zeta \left( 3 \right)}}{{8\pi}}
  \left(\frac{1}{\left(R-a_1\right)^2}+\frac{1}{\left(R+a_2\right)^2} \right)
  \nonumber\\ \vspace{-2cm}\hspace{-0cm} \dd-\frac{{\hbar cL\pi ^2 \left(R+a_2\right) \left(R-a_1\right) }}{{360 }}\left(\frac{1}{\left(R-a_1\right)^4}  +\frac{1}{\left(R+a_2\right)^4}\right)
  \nonumber\\ \vspace{1cm}\hspace{-0cm} \dd- \frac{{\hbar cL}}{4\pi}\sum\limits_{j = 1}^\infty
  {\frac{{\left( {e^{\frac{{2\pi j\left(R+a_2\right) }}{{\left(R-a_1\right) }}}  - 1} \right)\left(R-a_1\right)  +
  2\pi j\left(R+a_2\right) e^{\frac{{2\pi j\left(R+a_2\right) }}{{\left(R-a_1\right) }}} }}{{ \left(R-a_1\right) \left(R+a_2\right)^2 j^3\left( {e^{\frac{{2\pi j\left(R+a_2\right) }}{{\left(R-a_1\right) }}}  - 1}  \right)^2 }}}
  \nonumber\\ \hspace{-0cm} \dd- \frac{{\hbar cL}}{4\pi}\sum\limits_{j = 1}^\infty
  {\frac{{\left( {e^{\frac{{2\pi j\left(R-a_1\right) }}{{\left(R+a_2\right) }}}
  - 1} \right)\left(R+a_2\right)  +  2\pi j\left(R-a_1\right) e^{\frac{{2\pi j\left(R-a_1\right) }}{{\left(R+a_2\right) }}} }}{{ \left(R+a_2\right) \left(R-a_1\right)^2 j^3  \left( {e^{\frac{{2\pi j\left(R-a_1\right) }}{{\left(R+a_2\right) }}}  - 1} \right)^2 }}}
  \nonumber \\ +\left[a_1\to a_2,a_2\to a_1\right]\Bigg\} -\Bigg\{a_1\to b_1,a_2\to b_2\Bigg\}.
\end{eqnarray}
A Taylor expansion of the above expression up  to $\mathcal{O}\left( {\frac{1}{R}} \right)^4$ gives,
\begin{eqnarray}\label{ea3-2}
 \hspace{-1.1cm}\dd E_{{\rm{Cas.}}}^{{\rm{Outer}}}=\dd 2\Bigg\{- \frac{{\zeta \left( 3 \right)}}{{4\pi R^2 }} + \frac{{\zeta \left( 3 \right)}}{{4\pi R^3 }}\left( { - a_1  + a_2 } \right)-\frac{{  \pi ^2 }}{{180R^2 }} + \frac{{\pi ^2 }}{{180R^3 }}\left( { - a_1  + a_2 } \right)  \\ \vspace{1cm} \dd
 +\sum\limits_{j = 1}^\infty  {\left[ {\frac{{1 - e^{2\pi j} \left( {1 + 2\pi j} \right)}}{{2\pi j^3 \left( {e^{2\pi j}  - 1} \right)^2 R^2 }}} \right.}  + \left. {\frac{{\left( { - 1 + e^{2\pi j} \left( {1 + 2\pi j} \right)} \right)}}{{2\pi j^3 \left( {e^{2\pi j}  - 1} \right)^2 R^3 }}\left( { - a_1  + a_2 } \right)} \right] + \mathcal{O}\left( {\frac{1}{R}} \right)^4 \nonumber \\ +\left[a_1\to a_2,a_2\to a_1\right]\Bigg\} -\Bigg\{a_1\to b_1,a_2\to b_2\Bigg\}
 \end{eqnarray}
Note that the odd terms in $R$ cancel each other out. The even terms go to zero in the limit $R \rightarrow \infty$.

\section*{Acknowledgement} We would like to thank the research office
of the Shahid Beheshti University for financial support.

 \end{document}